%% file: main.tex
\documentclass[final,5p,times,twocolumn,numbers]{elsarticle}
\usepackage{amssymb,amsmath}
\usepackage{graphicx}
\usepackage{hyperref}
\usepackage{lipsum}
\biboptions{numbers,sort&compress}
\usepackage{makecell}
\newcommand{\BESIIIorcid}[1]{\href{https://orcid.org/#1}{\hspace*{0.1em}\raisebox{-0.45ex}{\includegraphics[width=1em]{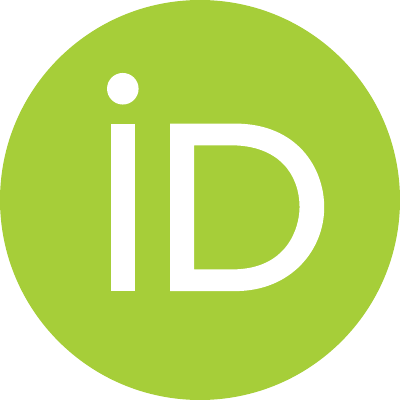}}}} 

\journal{Physics Letters B}

\begin{document}

\begin{frontmatter}
\title{First measurement of $\psi(3686)\to\bar{p} K^{+} \Sigma^{*}(1385)^{0}+c.c.$ and $\psi(3686)\to\bar{p} K^{*}(892)^{0} \Sigma^{+} + c.c.$}

\author{The BESIII Collaboration}

\begin{abstract}
Based on a sample of $(2712.4 \pm 14.3) \times 10^6$ $\psi(3686)$ events collected with the BESIII detector at the BEPCII collider, we report the first observation of the decays $\psi(3686)\to\bar{p} K^{+} \Sigma^{*}(1385)^{0}+c.c.$ and $\psi(3686)\to\bar{p} K^{*}(892)^{0} \Sigma^{+} + c.c.$. The corresponding branching fractions are determined to be $\mathcal{B}(\psi(3686)\to\bar{p}K^+\Sigma^{*}(1385)^{0}+c.c.)=(3.94\pm0.15_{\rm stat.}\pm0.24_{\rm syst.})\times10^{-5}$ and $\mathcal{B}(\psi(3686)\to\bar{p}K^{*}(892)^{0}\Sigma^++c.c.)=(4.93\pm0.14_{\rm stat.}\pm0.30_{\rm syst.})\times10^{-5}$. Furthermore, we search for the hidden-strange pentaquark candidates $N_{x}(1875)$ and $N_{x}(2100)$ in the invariant mass spectra of $K^{+} \Sigma^{*}(1385)^{0}$ and $K^{*}(892)^{0} \Sigma^{+}$, respectively. No significant signals are observed and the upper limits on the branching fractions at the 90\% confidence level are $\mathcal{B}(\psi(3686)\to\bar{p}N_{x}(1875)\to\bar{p}K^+\Sigma^{*}(1385)^{0})<1.58\times10^{-6}$ and $\mathcal{B}(\psi(3686)\to\bar{p}N_{x}(2100)\to\bar{p}K^{*}(892)^{0}\Sigma^+)<3.09\times10^{-7}$.
\end{abstract}


\end{frontmatter}

\section{Introduction}
In the conventional quark model, hadrons consist of either two or
three quarks ($q\bar{q}$ and $qqq$), whereas the quantum
chromodynamics allows for the existence of colour-singlet hadrons with
a larger number of (anti-)quarks, called exotic hadrons. The
exploration of such exotic hadrons is one of the most intriguing
topics in contemporary hadron physics. In 2015, the LHCb collaboration
reported the evidence for two hidden-charm pentaquarks, $P_c(4380)$ and $P_c(4450)$, in the $J/\psi p$ invariant mass spectrum of $\Lambda_b\to J/\psi pK$ decays~\cite{LHCb:2015yax}. These observations triggered extensive discussions on the nature of hidden-charm pentaquarks and motivated numerous theoretical proposals and experimental searches for additional pentaquark states~\cite{Chen:2016otp, He:2015cea, Chen:2019asm, LHCb:2019kea, Chen:2016ryt, Chen:2022onm, Liu:2005pm, BGOOD:2025cwg, Ramos:2013wua,Wu:2010jy}.

Because the masses of $P_c(4380)$ and $P_c(4450)$ lie close to the
$\Sigma_c^*D$ and $\Sigma_c D^*$ thresholds respectively, these $P_c$
states are naturally interpreted as molecular systems composed of
$\Sigma_c$ (or $\Sigma_c^*$) and $D$ (or $D^*$)~\cite{He:2015cea,
  Chen:2019asm}. Analogous to the interaction between $\Sigma_c$ (or
$\Sigma_c^*$) and $D$ (or $D^*$), the interaction between $\Sigma$ (or
$\Sigma^*$) and $K$ (or $K^*$) may be sufficiently attractive to
generate resonant structures. Consequently, pentaquarks analogous to
the $P_c$ states but with $s\bar{s}$ instead of $c\bar{c}$ may
exist. Some theoretical studies suggest that the baryonic resonance
$N_{x}(1875)$ with quantum numbers $I(J^P) = 1/2(3/2^-)$ generated by
$\Sigma^* K$ interaction, can be regarded as the hidden-strange
partner of $P_c(4380)$. Likewise, the baryonic resonance $N_{x}(2100)$
with the same quantum numbers $I(J^P) = 1/2(3/2^-)$ arising from
$\Sigma K^*$ interaction, can be viewed as the hidden-strange partner
of
$P_c(4450)$~\cite{Huang:2018ehi,Suo:2025rty,Wu:2023ywu,He:2017aps}. The
subscript $x$ is used to distinguish these states from the usual excited
nucleon states $N(1875)$ and $N(2100)$.  Despite the predictions,
direct observations of these resonance states in the hadronic decays
of charmonia have not been reported.

In this paper, using a sample of $(2712.4 \pm 14.3) \times 10^6$ $\psi(3686)$ events~\cite{BESIII:2024lks} recorded by the BESIII detector~\cite{BESIII:2009fln}, we analyse the decays $\psi(3686)\to\bar{p} K^{+} \Sigma^{*}(1385)^{0}$ and $\psi(3686)\to\bar{p} K^{*}(892)^{0} \Sigma^{+}$. Throughout this paper, $\Sigma^{*0}$ denotes $\Sigma^{*}(1385)^{0}$, $K^{*0}$ denotes $K^{*}(892)^{0}$, and charge-conjugate processes are implied. We report the first observation of these two decays and measure their branching fractions (BFs). Furthermore, we search for the hidden-strange pentaquark candidates $N_{x}(1875)$ and $N_{x}(2100)$ in the invariant mass spectra of $K^{+}\Sigma^{*0}$ and $K^{*0}\Sigma^{+}$, respectively. Since no significant resonant structures are observed, the upper limits on the BFs for $\psi(3686)\to\bar{p}N_{x}(1875)\to\bar{p} K^{+} \Sigma^{*0}$ and $\psi(3686)\to\bar{p}N_{x}(2100)\to\bar{p} K^{*0} \Sigma^{+}$ decays are determined at the 90\% confidence level (C.L.).

\section{BESIII Detector and Monte Carlo Simulation}

The BESIII detector~\cite{BESIII:2009fln} records symmetric $e^+e^-$ collisions 
provided by the BEPCII storage ring~\cite{Yu:2016cof}
in the center-of-mass energy range from 1.84 to 4.95~GeV, with a peak luminosity of $1.1 \times 10^{33}\;\text{cm}^{-2}\text{s}^{-1}$ 
achieved at $\sqrt{s} = 3.773\;\text{GeV}$. 
BESIII has collected large data samples in this energy region~\cite{BESIII:2020nme, Lu:2020imt, Zhang:2022bdc}. The cylindrical core of the BESIII detector covers 93\% of the full solid angle and consists of a helium-based
 multilayer drift chamber~(MDC), a time-of-flight
system~(TOF), and a CsI(Tl) electromagnetic calorimeter~(EMC),
which are all enclosed in a superconducting solenoidal magnet
providing a 1.0~T magnetic field. The solenoid is supported by an
octagonal flux-return yoke with resistive plate counter muon
identification modules interleaved with steel. 

The charged-particle momentum resolution at $1~{\rm GeV}/c$ is
$0.5\%$, and the 
${\rm d}E/{\rm d}x$
resolution is $6\%$ for electrons
from Bhabha scattering. The EMC measures photon energies with a
resolution of $2.5\%$ ($5\%$) at $1$~GeV in the barrel (end cap)
region. The time resolution in the plastic scintillator TOF barrel region is 68~ps, while that in the end cap region was 110~ps. The end cap TOF system was upgraded in 2015 using multigap resistive plate chamber
technology, providing a time resolution of
60~ps,
which benefits 83\% of the data used in this analysis~\cite{Li:2017jpg, Guo:2017sjt, Cao:2020ibk}.

Monte Carlo (MC) simulated data samples produced with a {\sc
geant4}-based~\cite{GEANT4:2002zbu} software package, which
includes the geometric description of the BESIII detector and the
detector response, are used to determine detection efficiencies
and to estimate backgrounds. The simulation models the beam
energy spread and initial state radiation (ISR) in the $e^+e^-$
annihilations with the generator {\sc
kkmc}~\cite{Jadach:1999vf, Jadach:2000ir}. The inclusive MC sample includes the production of the
$\psi(3686)$ resonance, the ISR production of the $J/\psi$, and
the continuum processes incorporated in {\sc
kkmc}~\cite{Jadach:1999vf, Jadach:2000ir}. All particle decays are modelled with {\sc
evtgen}~\cite{Lange:2001uf,Ping:2008zz} using BFs 
either taken from the
Particle Data Group (PDG)~\cite{ParticleDataGroup:2024cfk}, when available,
or otherwise estimated with {\sc lundcharm}~\cite{Chen:2000tv,Yang:2014vra}. Final state radiation
from charged final state particles is incorporated using the {\sc
photos} package~\cite{Barberio:1990ms}. To optimize the selection criteria, samples of 1 million simulated events are generated for each decay channel according to the phase space (PHSP) model. Additionally, samples of 1 million simulated events from partial-wave analysis (PWA) are also generated for each decay channel to determine the detection efficiency.

\section{Event Selection}
\label{sec:selection}

For the decay $\psi(3686) \to \bar{p} K^+ \Sigma^{*0}$ with $\Sigma^{*0} \to \Lambda \pi^0$, the $\Lambda$ and $\pi^0$ are reconstructed via $p\pi^-$ and $\gamma\gamma$, respectively. For the decay $\psi(3686) \to \bar{p} K^{*0} \Sigma^+$ with $K^{*0} \to K^+ \pi^-$ and $\Sigma^+ \to p \pi^0$, the $\pi^0$ is also reconstructed via $\gamma\gamma$. These two decay processes share the same final-state particles.

Charged tracks detected in the MDC are required to
be within a polar angle ($\theta$) range of $|\cos\theta|<0.93$, where $\theta$ is defined with respect to the $z$-axis, which is the symmetry axis of the MDC. For charged tracks originating from $\Lambda$ and $\Sigma^+$ decays, the distance of closest approach to the interaction point must be less than 30\,cm along the $z$-axis, $|V_{z}|$,  and less than 15\,cm in the transverse plane, $|V_{xy}|$; for other charged tracks, the corresponding limits are 10\,cm and 1\,cm, respectively.

Particle identification (PID) for charged tracks combines measurements of the energy deposited in the MDC~(d$E$/d$x$) and the flight time in the TOF to form likelihoods $\mathcal{L}(h)$ $(h=p, K, \pi)$ for each hadron ($h$) hypothesis. Each track  is identified as the particle  type with the greatest likelihood. Events containing one kaon, one pion, one proton, and one antiproton, with zero total charge, are retained for further analysis.

Photon candidates are identified using isolated showers in the EMC.  The deposited energy of each shower must be more than 25~MeV in the barrel region ($|\cos \theta|< 0.80$) and more than 50~MeV in the end cap region ($0.86 <|\cos \theta|< 0.92$). To exclude showers that originate from charged tracks, the angle subtended by the EMC shower and the position of the closest charged track at the EMC must be greater than 5 degrees as measured from the interaction point. To suppress electronic noise and showers unrelated to the event, the difference between the EMC time and the event start time is required to be within [0, 700]\,ns. The number of photon candidates is required to be at least two per event.

To  reconstruct the $\pi^0$ candidates,  the invariant mass of $\gamma\gamma$ is constrained to the nominal mass of $\pi^0$ from the PDG~\cite{ParticleDataGroup:2024cfk} via a one-constraint (1C) kinematic fit. The combinations satisfying $\chi^2_{1\rm{C}} < 10$ are retained.

To reconstruct the $\Lambda$ candidate from the process $\psi(3686) \to \bar{p} K^+ \Sigma^{*0}$ with $\Sigma^{*0} \to \Lambda\pi^0$, a secondary vertex fit is applied to the $p\pi^-$ combination. To suppress potential backgrounds, the combination satisfying $|M(p\pi^-) - m_{\Lambda}| < 0.007$ GeV/$c^{2}$ is selected as the $\Lambda$ candidate, where $M(p\pi^-)$ is the proton-pion invariant mass and $m_{\Lambda}$ is the nominal mass of $\Lambda$ from the PDG~\cite{ParticleDataGroup:2024cfk}. 

To suppress backgrounds and improve mass resolution, a five-constraint
(5C) kinematic fit is performed under the hypotheses of $\psi(3686)
\to \bar{p}K^+\Lambda\pi^{0}$ and $\psi(3686) \to
\bar{p}K^+\pi^{-}p\pi^{0}$ decays, respectively, including
energy-momentum conservation and the invariant mass constraint for
$\pi^0$. The optimal $\pi^0$ candidate is selected with the minimum
chis-square statistic of the 5C kinematic fit $\chi^2_{5\rm{C}}$. The resulting $\chi^2_{5\rm{C}}$ are required to be less than 45 and 200 for the processes $\psi(3686)\to\bar{p} K^{+} \Sigma^{*0}$ and $\psi(3686)\to\bar{p} K^{*0} \Sigma^{+}$, respectively, as determined by optimising the figure of merit $\frac{S}{\sqrt{S+B}}$, where $S$ is the normalised number of signal events in the signal MC sample and $(S+B)$ corresponds to the number of data events.

\section{Measurement of branching fractions}

\subsection{Analysis of $\psi(3686) \to \bar{p} K^+ \Sigma^{*0}$}

\begin{figure}
\centering
\includegraphics[width=0.33\textwidth]{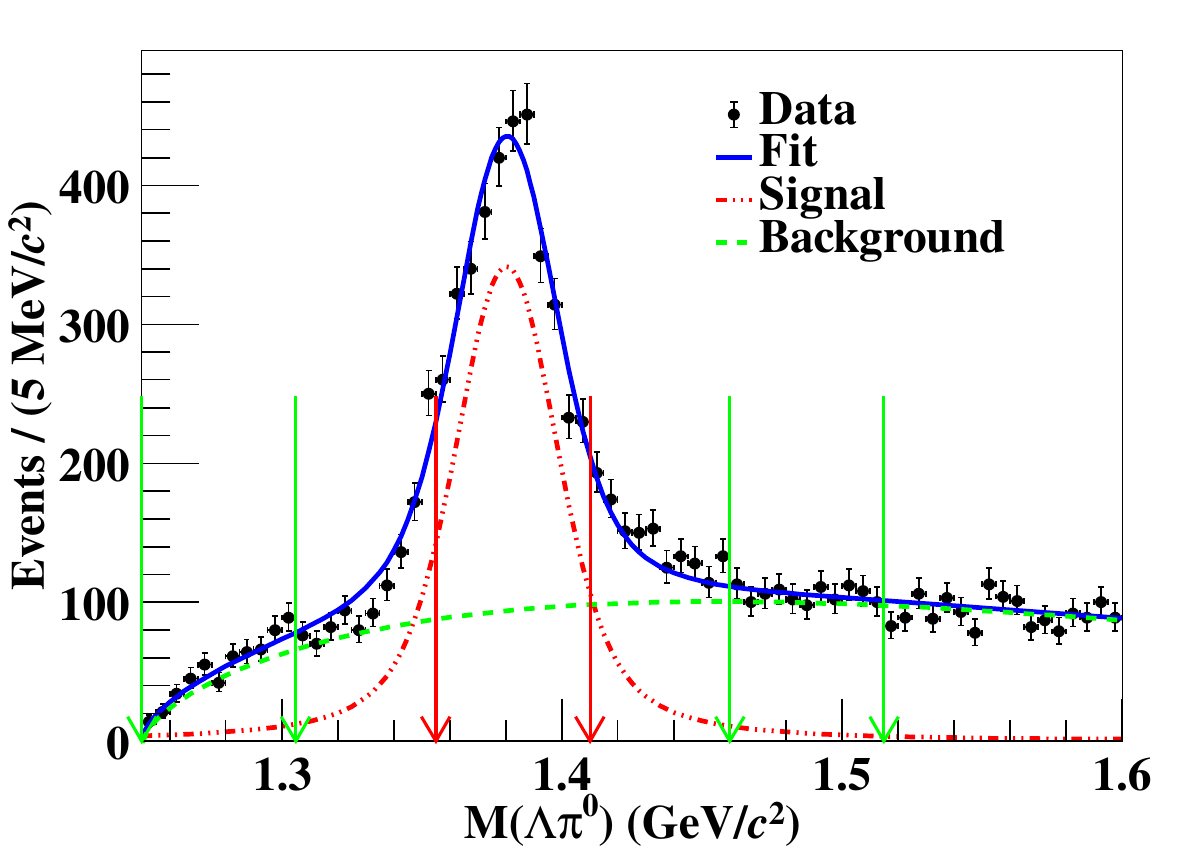}
   \caption{Invariant mass distribution of $\Lambda\pi^0$. The red and green solid arrows indicate the $\Sigma^{*0}$ signal and sideband regions, respectively.}\label{fig_sideband}
\end{figure}

After the selection criteria are applied, a clear $\Sigma^{*0}$ peak is observed in the invariant mass spectrum of $\Lambda\pi^0$ shown in Fig.~\ref{fig_sideband}. Potential background sources are  investigated using an inclusive MC sample of 2.7 billion $\psi(3686)$ events, and no remaining background events are found to contain $\Sigma^{*0}$. 

To determine the number of signal events $N_{\rm{sig}}$, an unbinned
maximum-likelihood fit is performed to the $\Lambda\pi^0$ invariant
mass spectrum, with interference effects neglected. The $\Sigma^{*0}$
signal shape is described by a Breit-Wigner (BW) function convolved
with a Gaussian function to account for the detector resolution. The
mass and width of the BW function are free in the fit. The Gaussian
parameters are obtained by fitting the deviation distribution of
the reconstructed and the true masses in the signal MC events. An ARGUS~\cite{ARGUS:1990hfq} function is used to describe the background shape, with its threshold parameter fixed and all other parameters floating. Figure~\ref{fig_sideband} shows the fit result, yielding $4473\pm173$ signal events in the fit range.
The mass and width of the BW function are determined to be  $1.382\pm0.001\ \text{GeV}/c^2$
 and 
$0.041\pm0.002\ \text{GeV}/c^2$
, respectively.

\begin{figure*}[tb]
\centering
\includegraphics[width=0.33\textwidth]{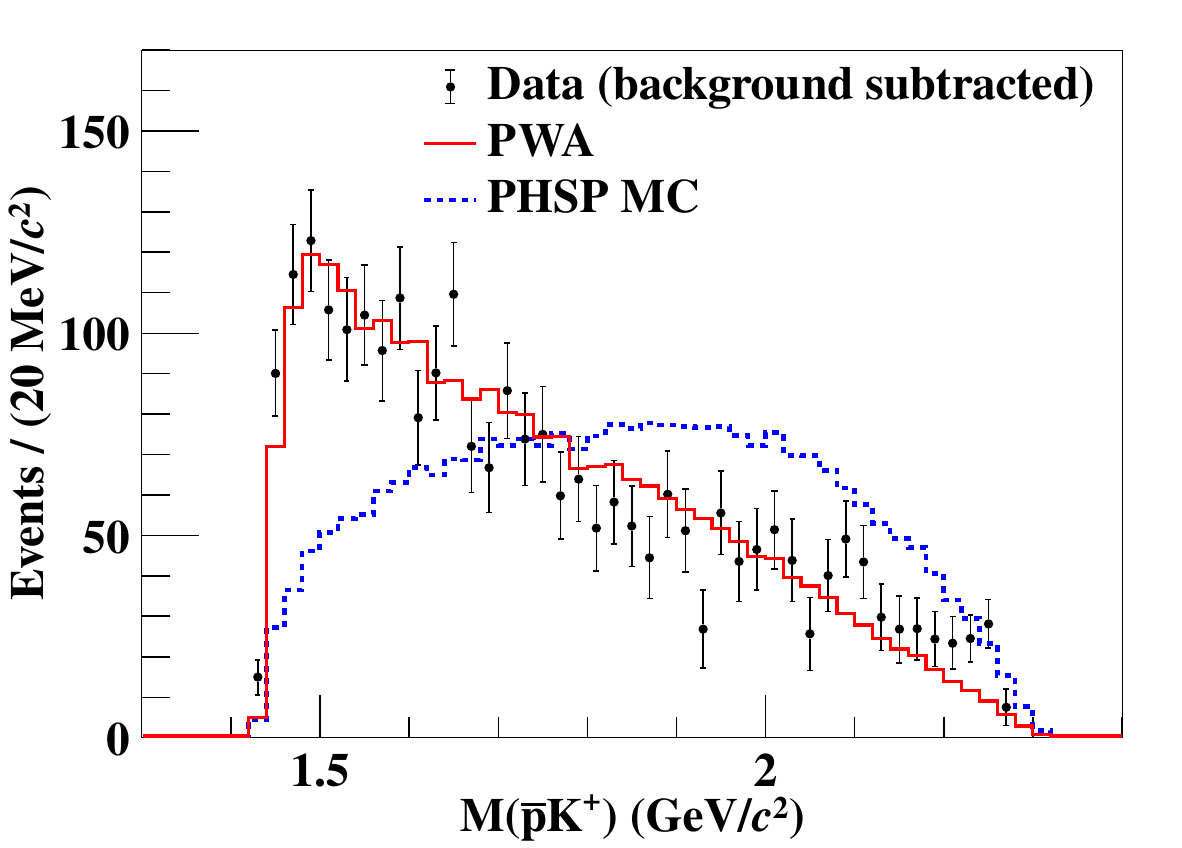}
\includegraphics[width=0.33\textwidth]{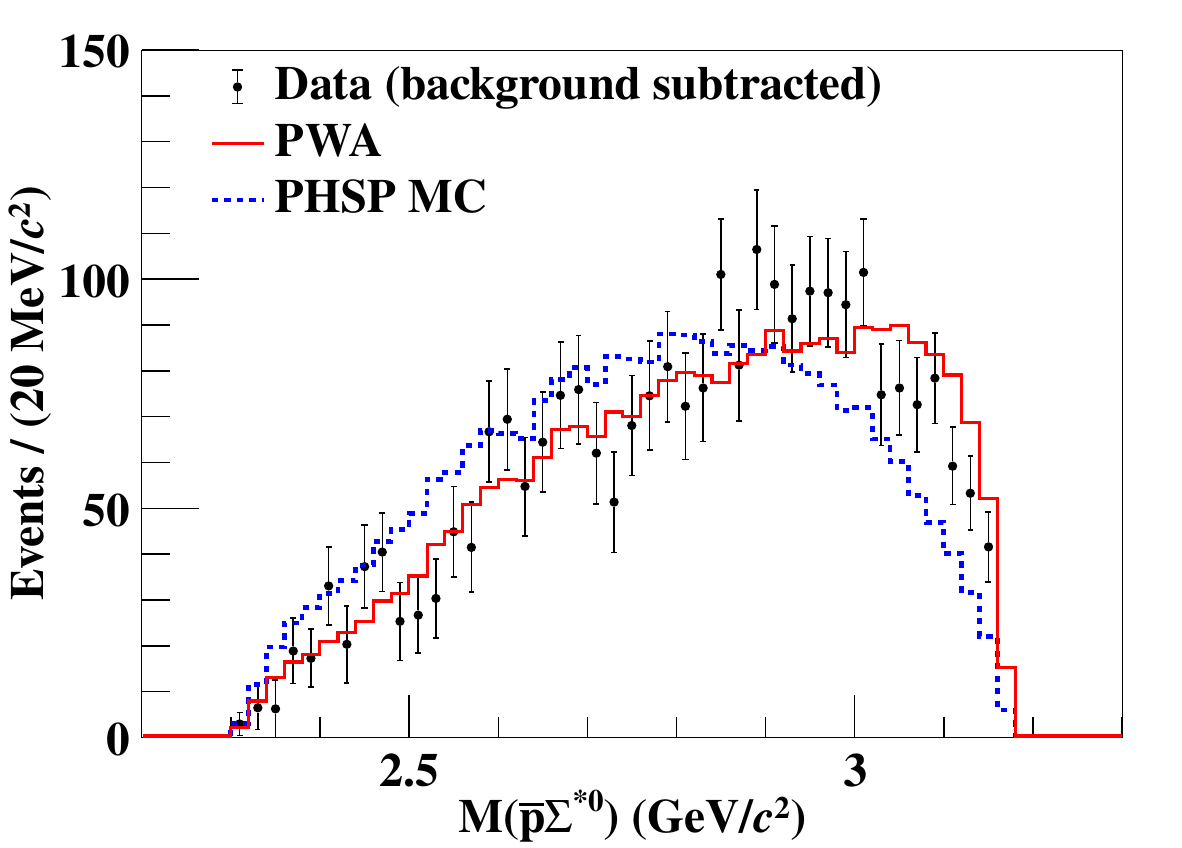} 
\includegraphics[width=0.33\textwidth]{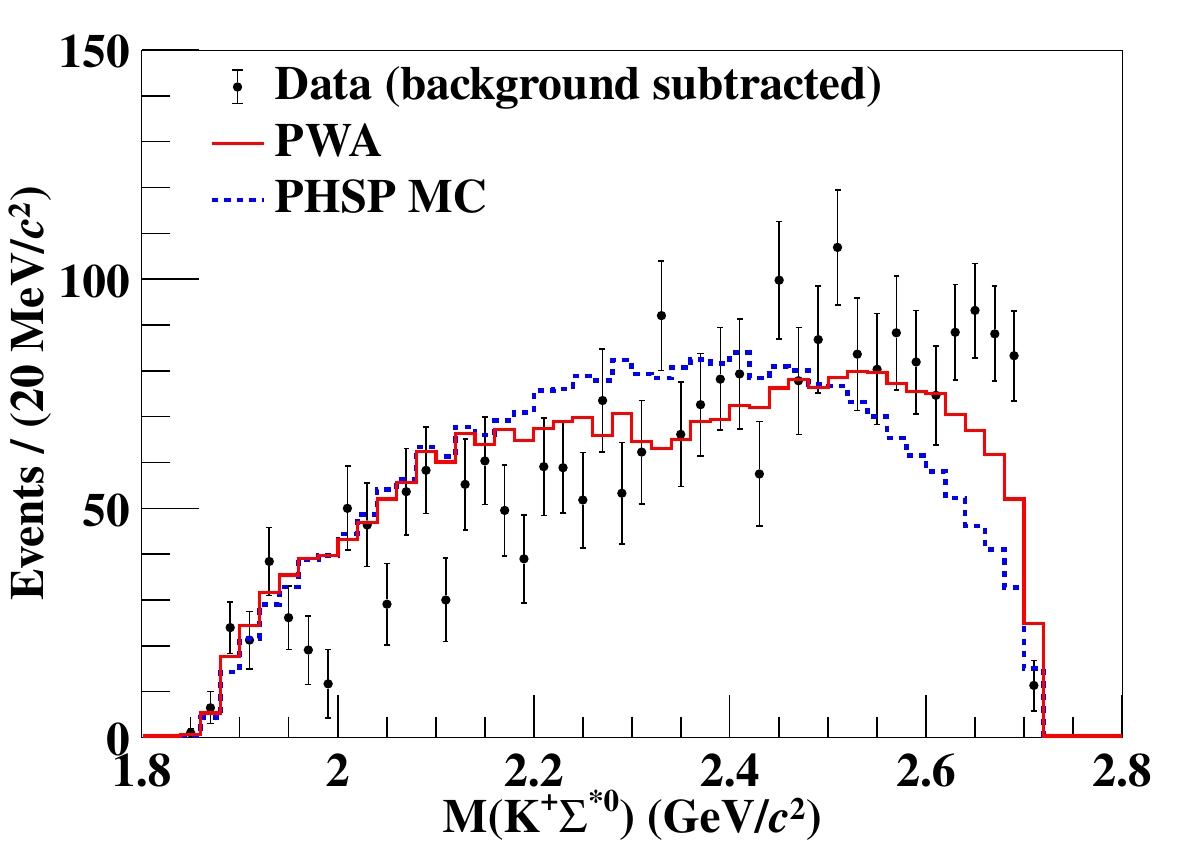}
\caption{Comparisons between data and PWA projections for the invariant mass distributions of  $\bar{p}K^{+}$, $\bar{p}\Sigma^{*0}$ and $K^{+}\Sigma^{*0}$ from $\psi(3686) \to \bar{p} K^{+} \Sigma^{*0}$. Points with error bars are data with background subtracted, where the background is estimated from the corrected sideband. The red histograms are PWA projections and the blue histograms are PHSP MC.}
\label{pwa1}
\end{figure*}

The $\bar{p} K^+$, $\bar{p} \Sigma^{*0}$, and  $K^+ \Sigma^{*0}$ invariant mass spectra of Fig.~\ref{pwa1} show a discrepancy between the data and the PHSP MC sample. To better determine the efficiency, a minimal PWA is performed using the TFPWA framework~\cite{Jiang:2024vbw}, based on an unbinned maximum-likelihood fit. In the log-likelihood calculation, we use $\Sigma^{*0}$ sideband events to estimate the background contributions. The $\Sigma^{*0}$ signal region is $[1.355,1.410]\ \text{GeV}/c^2$, with sideband regions $[1.250,1.305]\ \text{GeV}/c^2$ and $[1.460,1.515]\ \text{GeV}/c^2$. The log-likelihood is calculated as $\ln L_{\rm signal} - \sum \omega \cdot\ln L_{\rm sideband}$, where $L_{\rm signal}$ and $L_{\rm sideband}$ are the likelihood values of events in the $\Sigma^{*0}$ signal and sideband regions, respectively, and the weighting factor $\omega$ is fixed according to the sideband normalization. The PWA fit includes the intermediate resonance $\Lambda(1405)$ and a non-resonant contribution with $J^P=3/2^-$. The PWA result is used to generate MC sample with improved consistency relative to data. Figure~\ref{pwa1} shows the invariant masses: $M(\bar{p}K^{+})$, $M(\bar{p}\Sigma^{*0})$ and $M(K^{+}\Sigma^{*0})$ from data, PWA result, and PHSP MC sample. The resulting MC sample from the PWA fit provides a better description of the data than the PHSP sample, and is employed for the efficiency determination.

The BF of the decay $\psi(3686) \to \bar{p} K^+ \Sigma^{*0}$ is calculated by
\begin{equation}
\mathcal{B}(\psi(3686) \to \bar{p} K^+ \Sigma^{*0}) = \frac{N_{\text{sig}}}{N_{\psi(3686)} \cdot \prod_{i} \mathcal{B}_{i} \cdot \epsilon} ,
\end{equation}
where $N_{\text{sig}}=4473\pm173$ is the number of $\Sigma^{*0}$ signal events obtained from the aforementioned fit; $N_{\psi(3686)}=(2712.4 \pm 14.3) \times 10^6$ is the total number of $\psi(3686)$ events~\cite{BESIII:2024lks}; $\prod_{i} \mathcal{B}_{i}=\mathcal{B}(\Sigma^{*0} \to \Lambda \pi^0) \cdot \mathcal{B}(\Lambda \to p \pi^-) \cdot \mathcal{B}(\pi^0 \to \gamma\gamma)$, where $\mathcal{B}_{i}$ are results from the PDG~\cite{ParticleDataGroup:2024cfk}; the efficiency $\epsilon=(7.63 \pm 0.03)\%$ is estimated with MC sample generated according to the PWA result. The BF for the decay $\psi(3686) \to \bar{p} K^+ \Sigma^{*0}$ is then determined to be $(3.94 \pm 0.15 ) \times 10^{-5}$, where the uncertainty is only statistical.

\subsection{Analysis of $\psi(3686) \to \bar{p} K^{*0} \Sigma^{+}$}

\begin{figure*}[tb]
\centering
\includegraphics[width=0.33\textwidth]{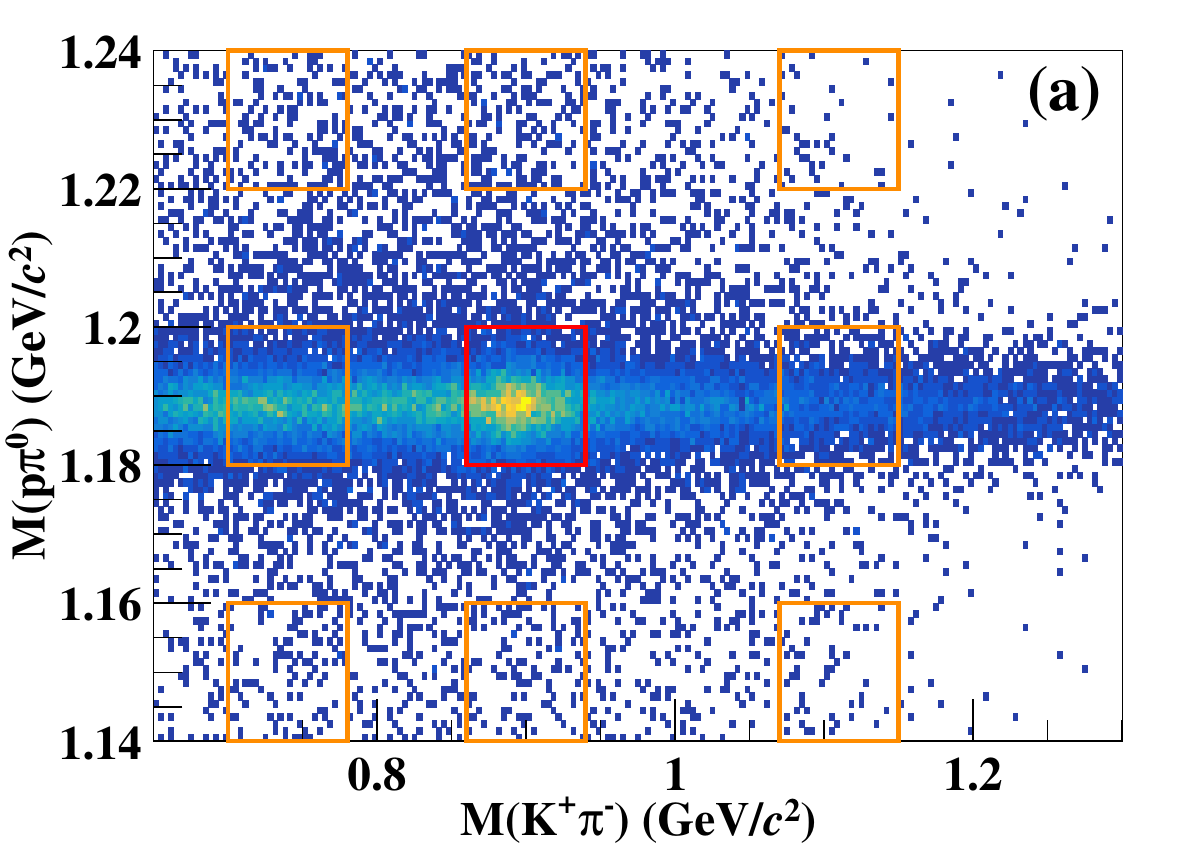}
\includegraphics[width=0.33\textwidth]{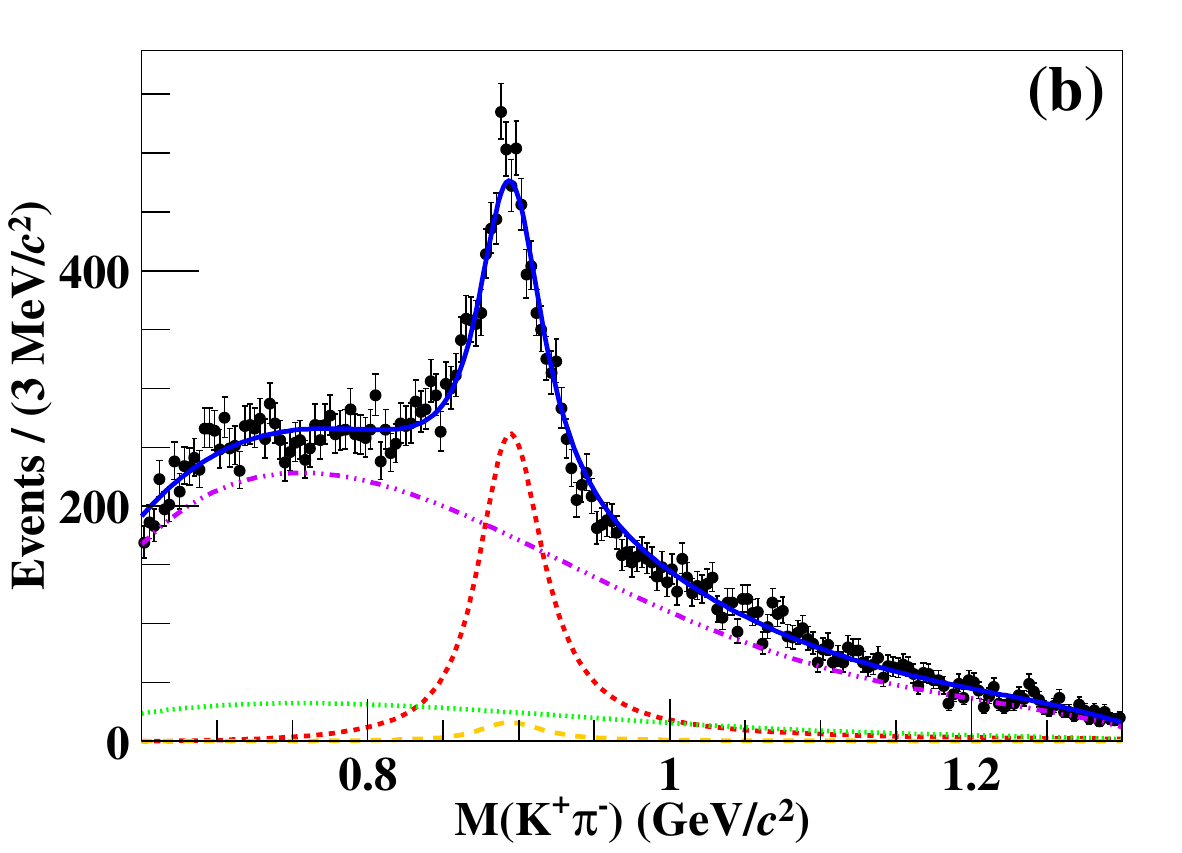}
\includegraphics[width=0.33\textwidth]{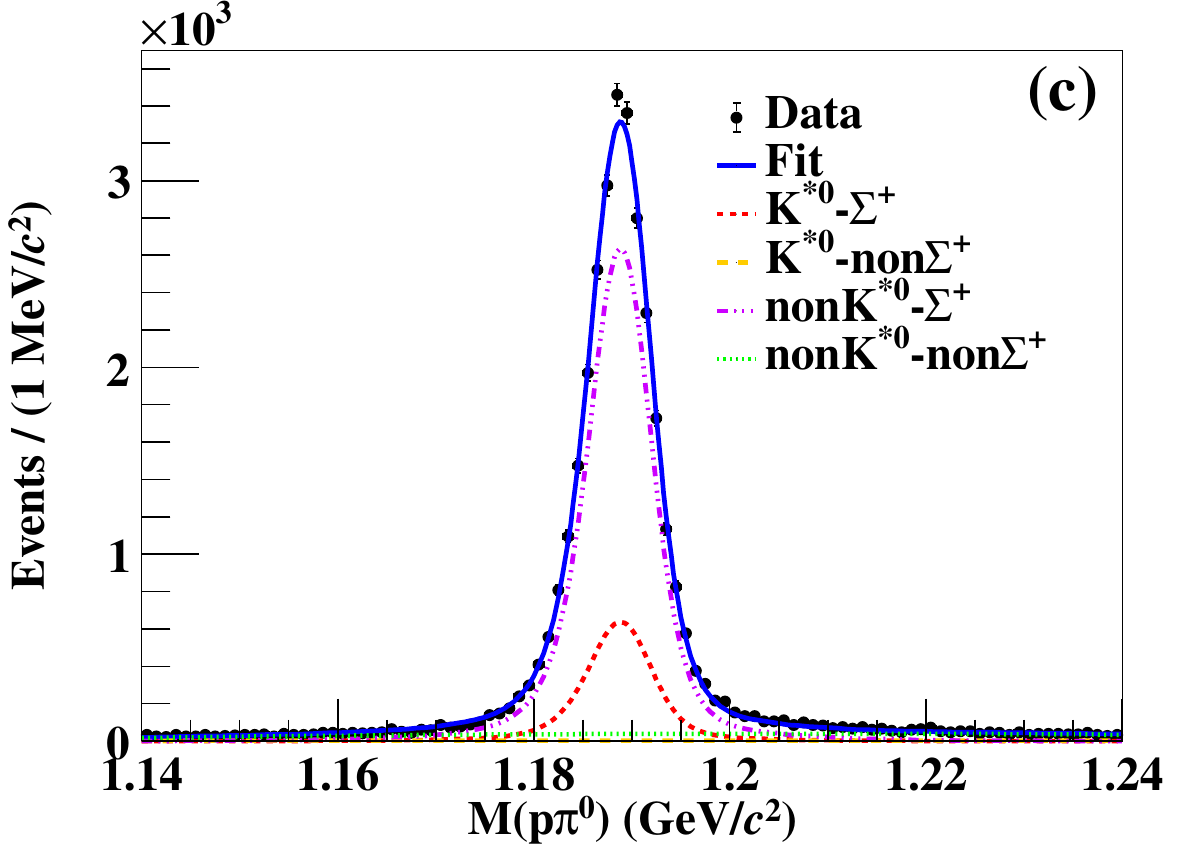}
    \caption{(a) Two-dimensional invariant mass spectrum of $K^{*0}$–$\Sigma^{+}$ with signal and sideband regions, where the red box denotes the signal region and the orange boxes denote the sideband regions; (b) $M(K^+\pi^-)$ and (c) $M(p\pi^0)$ projections of the $\psi(3686) \to \bar{p} K^{*0} \Sigma^+$ data.}\label{fig_2dsideband}
\end{figure*}

After the selection criteria are applied, clear $K^{*0}$ and $\Sigma^{+}$ peaks are observed in the invariant mass spectra of $K^{+}\pi^-$ and $p\pi^0$, respectively, as shown in Fig.~\ref{fig_2dsideband}(b) and Fig.~\ref{fig_2dsideband}(c). Potential background sources are  investigated using an inclusive MC sample of 2.7 billion $\psi(3686)$ events, and none of the remaining background events are found to contain both $K^{*0}$ and $\Sigma^{+}$. 

To determine the number of signal events $N_{\rm{sig}}$, we perform an unbinned two-dimensional combined maximum-likelihood fit to $M(K^{+}\pi^-)$ and $M(p\pi^0)$ invariant mass spectra, with interference effects neglected. In the $K^+\pi^-$ channel, the $K^{*0}$ signal distribution is described by a P-wave BW function convolved with a double Gaussian function, whereas the non-$K^{*0}$ background is modelled by a fourth-order polynomial. The BW function's parameters are fixed to the PDG values~\cite{ParticleDataGroup:2024cfk}. The double Gaussian parameters are obtained by fitting the deviation distribution of reconstructed and true masses in signal MC sample. The polynomial's parameters are free in the fit. In the $p\pi^0$ channel, the $\Sigma^+$ signal is described by the signal MC shape convolved with a Gaussian function, and the non-$\Sigma^+$ background by a second-order polynomial. Both the Gaussian function and the polynomial's parameters are free in the fit. The $K^{*0}$ signal region is defined as $[0.86,0.94]\ \text{GeV}/c^2$, with sideband regions $[0.70,0.78]\ \text{GeV}/c^2$ and $[1.07,1.15]\ \text{GeV}/c^2$. The $\Sigma^+$ signal region is $[1.18,1.20]\ \text{GeV}/c^2$, with sideband regions $[1.14,1.16]\ \text{GeV}/c^2$ and $[1.22,1.24]\ \text{GeV}/c^2$. Figure~\ref{fig_2dsideband}(a) shows the signal and sideband regions, and the fit  yields $3734\pm103$ signal events in the signal region.

\begin{figure*}[tb]
    \centering
\includegraphics[width=0.33\textwidth]{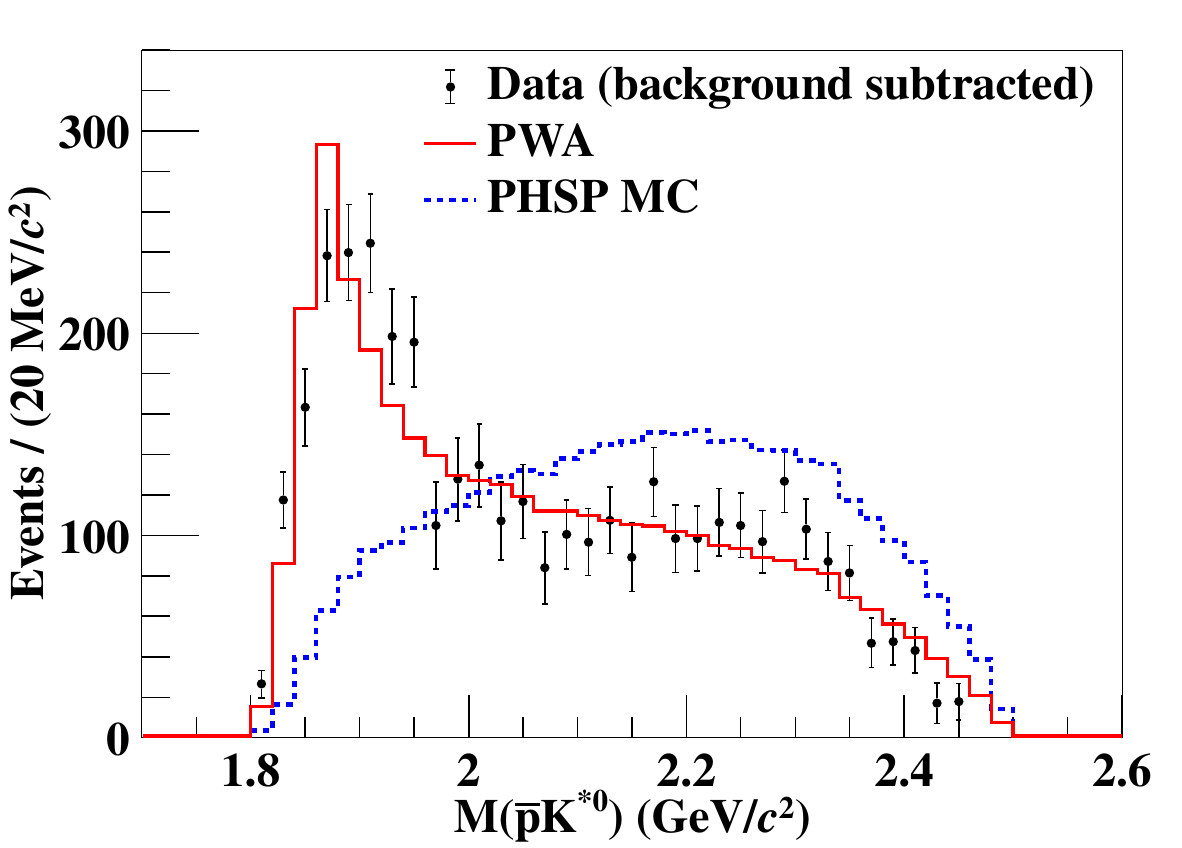}
\includegraphics[width=0.33\textwidth]{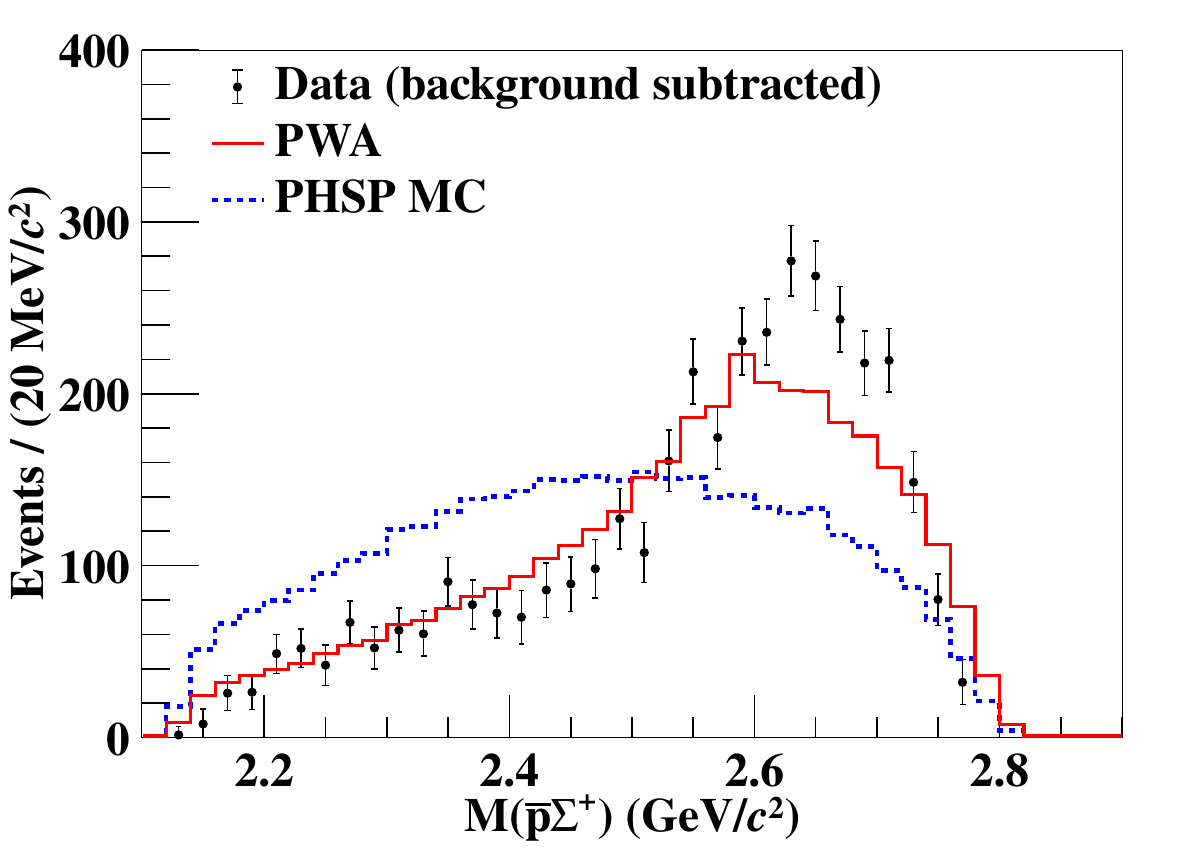}
\includegraphics[width=0.33\textwidth]{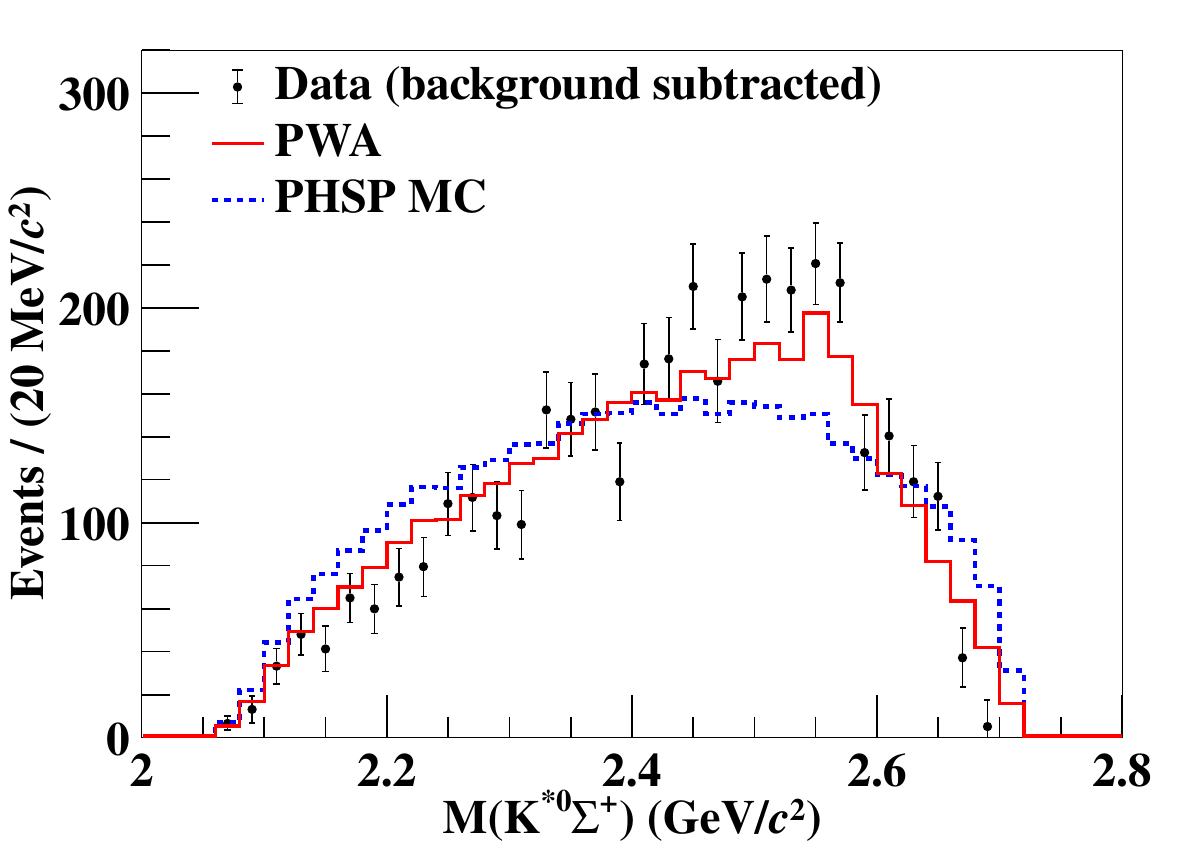}  
\caption{Comparisons between data and PWA projections for the invariant mass distributions of  $\bar{p}K^{*0}$, $\bar{p}\Sigma^{+}$ and $K^{*0}\Sigma^{+}$ from $\psi(3686) \to \bar{p} K^{*0} \Sigma^{+}$. Points with error bars are data with background subtracted, where the background is estimated from the corrected sideband. The red histograms are PWA projections and the blue histograms are PHSP MC.}
\label{pwa2}
\end{figure*}

A minimal PWA is performed for this channel, incorporating the
intermediate resonance $\Sigma(1880)$ and a non-resonant contribution
with $J^P=1/2^-$. Background contributions are estimated using the
$K^{*0}$--$\Sigma^{+}$ two-dimensional sideband. Figure~\ref{pwa2}
shows the invariant masses: $M(\bar{p}K^{*0})$, $M(\bar{p}\Sigma^{+})$
and $M(K^{*0}\Sigma^{+})$ from the data, the PWA fit, and the PHSP MC sample. The resulting MC sample from the PWA fit provides a better description of the data than the PHSP sample, and is employed for the efficiency determination.

The BF of the decay $\psi(3686) \to \bar{p} K^{*0} \Sigma^{+}$ is calculated by
\begin{equation}
\mathcal{B}(\psi(3686) \to \bar{p} K^{*0} \Sigma^{+}) = \frac{N_{\text{sig}}}{N_{\psi(3686)} \cdot \prod_{i} \mathcal{B}_{i} \cdot \epsilon} ,
\end{equation}
where $N_{\text{sig}}=3734\pm103$ is the number of signal events obtained from the fit; $N_{\psi(3686)}=(2712.4 \pm 14.3) \times 10^6$ is the total number of $\psi(3686)$ events~\cite{BESIII:2024lks}; $\prod_{i} \mathcal{B}_{i}=\mathcal{B}(K^{*0} \to K^+ \pi^-) \cdot \mathcal{B}(\Sigma^+ \to p \pi^0) \cdot \mathcal{B}(\pi^0 \to \gamma\gamma)$, where $\mathcal{B}_{i}$ are results from the PDG~\cite{ParticleDataGroup:2024cfk}; the efficiency $\epsilon=(8.28 \pm 0.03)\%$ is estimated with MC sample generated according to the PWA result. The BF for the decay $\psi(3686) \to \bar{p} K^{*0} \Sigma^{+}$ is then determined to be $(4.93 \pm 0.14 ) \times 10^{-5}$, where the uncertainty is only statistical.

\subsection{Systematic uncertainties}
\label{syserr}
Various sources of systematic uncertainties for $\mathcal{B}(\psi(3686) \to \bar{p} K^{+} \Sigma^{*0})$ and $\mathcal{B}(\psi(3686) \to \bar{p} K^{*0} \Sigma^{+})$ have been investigated and discussed in detail below. 

\begin{enumerate}
    \item Number of $\psi(3686)$ events. The number of $\psi(3686)$ events is determined by analyzing inclusive hadronic $\psi(3686)$ decays with an uncertainty of 0.5\% \cite{BESIII:2024lks}.

     \item Reconstruction of $\Lambda$ or $\Sigma^+$. The uncertainties associated with the $\Lambda$ and $\Sigma^+$ reconstruction are 2.0\% and 1.9\%, quoted from Ref.~\cite{BESIII:2023ojh} and Ref.~\cite{BESIII:2024umc}, respectively.
     
    \item MDC tracking and PID. The systematic uncertainty due to MDC tracking and PID is estimated with the $J/\psi \to p\bar{p}\pi^+\pi^-$~\cite{Yuan:2015wga} and $e^+e^- \to \pi^+\pi^-K^+K^-$~\cite{BESIII:2019gjc} control samples by studying the difference between data and MC simulation efficiencies. The data-MC difference for MDC tracking efficiency is 1.0\% per charged track, and that for PID efficiency is also 1.0\%. The charged tracks from the $\Lambda$ and $\Sigma^{+}$ decays are considered in the uncertainties from $\Lambda$ reconstruction and $\Sigma^{+}$ reconstruction, respectively.

    \item Photon detection. For photons detected by the EMC, the detection efficiency is studied using a control sample of $e^+e^-\to\gamma_{\text{ISR}}\mu^+\mu^-$~\cite{Prasad:2015bra}. The systematic uncertainty, defined as the relative difference in efficiencies between data and MC simulation, is observed to be 1.0\% per photon. For the decay $\psi(3686) \to \bar{p} K^{+} \Sigma^{*0}$, with two photons reconstructed, we take the systematic uncertainty to be 2.0\%. For the decay $\psi(3686) \to \bar{p} K^{*0} \Sigma^{+}$, the $\pi^{0}$ reconstruction is considered in the uncertainty from the $\Sigma^+$ reconstruction.

    \item Quoted BFs. The uncertainties on the intermediate decay BFs in the signal processes are taken from the PDG values \cite{ParticleDataGroup:2024cfk}.

    \item Kinematic fit. To investigate the systematic uncertainty associated with the kinematic fit, the track helix parameter correction method is used \cite{BESIII:2012mpj}. The difference in the detection efficiencies with and without the helix correction is taken as the systematic uncertainty.

    \item Signal and background shapes. For the two decay processes, the signal shapes are replaced by signal MC convolved with a Gaussian function, while all background shapes are replaced by second-order polynomials. The resultant changes in the BFs are taken as the systematic uncertainties.

    \item Fit range. The systematic uncertainties due to different fit ranges are examined by enlarging and shrinking the fit range eight times. For each case, the deviation between the alternative and nominal fits is defined as $\xi=\frac{|a_1-a_2|}{\sqrt{|\sigma_1^2-\sigma_2^2|}}$, where $a_1\pm \sigma_1$ and $a_2\pm\sigma_2$ are the nominal and alternative fit results, respectively.  If $\xi$ is less than 2.0, the associated systematic uncertainty is negligible according to the Barlow test \cite{Barlow:2002yb}. The largest difference is assigned as the systematic uncertainty.

    \item Detection efficiency. To evaluate the systematic uncertainty of the detection efficiency derived from the PWA, we generate an additional MC sample in which the intermediate resonance $\Lambda(1405)$ in the decay $\psi(3686) \to \bar{p} K^{+} \Sigma^{*0}$ is replaced by a resonance with quantum numbers $J^P = 1/2^+$, and $\Sigma(1880)$ in the decay $\psi(3686) \to \bar{p} K^{*0} \Sigma^{+}$ is replaced by $\Sigma(1940)$. The differences in detection efficiencies between these two distinct MC samples are assigned as the systematic uncertainties.
\end{enumerate}

Assuming all the sources of systematic uncertainty above are independent, the total systematic uncertainties for the two decay modes are obtained by adding them in quadrature. The results are summarised in Table~\ref{tab_sysBr}.

 \begin{table}
		\begin{center}
			\caption{Systematic uncertainties (in \%) on the measurements of the BFs.}\label{tab_sysBr}
			\begin{tabular}{c|c|c}\hline\hline
                       & ~~~~$\psi(3686) \to$~~~~& ~~~~$\psi(3686) \to$~~~~\\
				Source& ~~~~$\bar{p} K^+ \Sigma^{*0}$~~~~& ~~~~$\bar{p} K^{*0} \Sigma^+$~~~~\\\hline
             $N_{\psi(3686)}$ & 0.5 &0.5\\
             $\Lambda$ reconstruction &2.0 & -\\
             $\Sigma^+$ reconstruction & - & 1.9\\
		   MDC tracking & 2.0 & 3.0 \\
          PID & 2.0&3.0\\
		   Photon detection & 2.0 & -\\
             Quoted BFs& 1.9 & 0.6 \\
             Kinematic fit &4.1&0.4\\
             Signal shape &0.3&0.5\\
             Background shape &1.1&0.2\\
            Fit range &$<0.1$&3.6\\
            Detection efficiency &0.4 & 0.6\\
            \hline
             Total &6.2& 6.0\\
             \hline\hline
			\end{tabular}
		\end{center}
		\end{table}

\section{Search for $N_{x}(1875)$ and $N_{x}(2100)$}
\subsection{Search for $\psi(3686) \to\bar{p}N_{x}(1875)\to \bar{p} K^{+} \Sigma^{*0}$}\label{sec:search1875}

\begin{figure}
\centering
\includegraphics[width=0.33\textwidth]{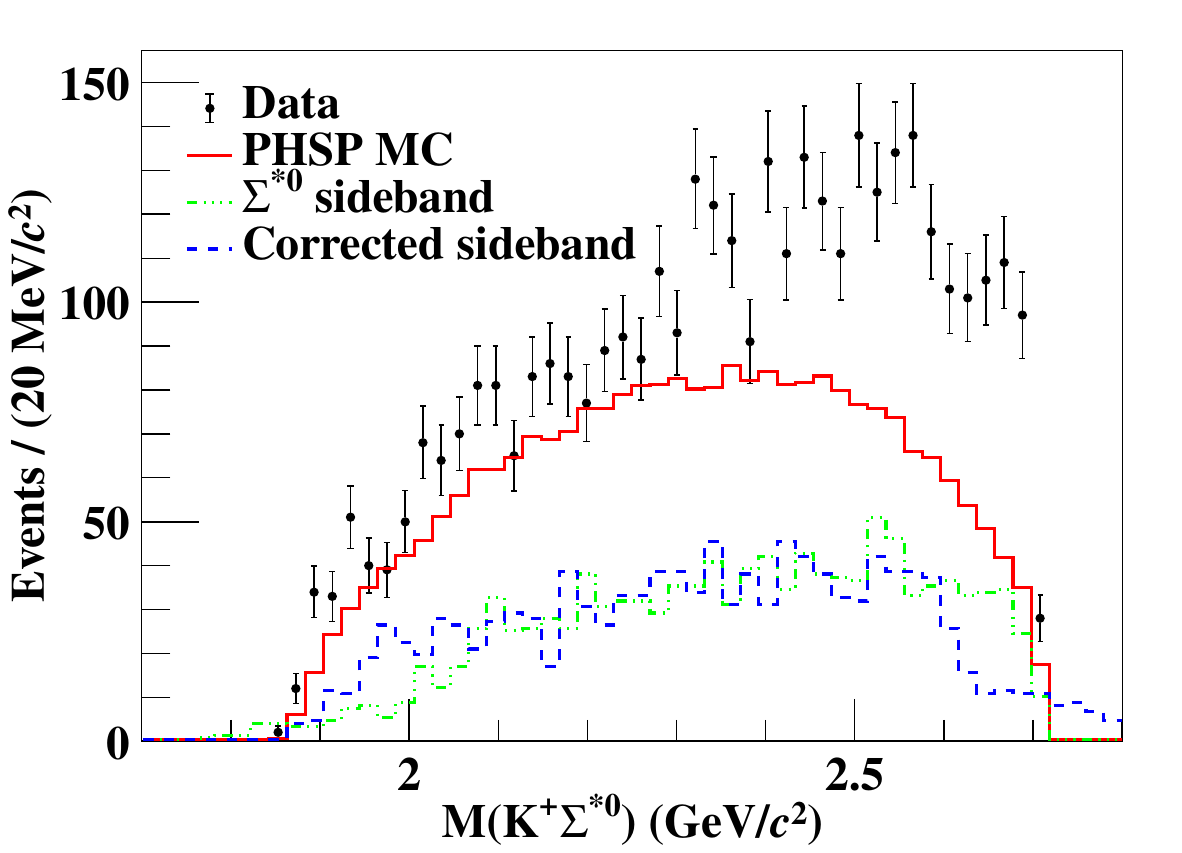}
    \caption{Invariant mass distribution of $K^+\Sigma^{*0}$. The dots
      with error bars are the data; the red solid, green dot-dashed and blue
      dashed lines represent the PHSP MC sample, $ \Sigma^{*0} $ sidebands and corrected sidebands, respectively. The PHSP MC is normalized to background-subtracted data events.}\label{fig_shift1}
\end{figure}

 Potential backgrounds are studied with inclusive MC samples of $\psi(3686)$ decays. The backgrounds are found to fall into two categories: non-$\Sigma^{*0}$ and non-resonant.
Non-$\Sigma^{*0}$ backgrounds are estimated using the $\Sigma^{*0}$ sidebands in the data. The kinematic regions of events in the $\Sigma^{*0}$ sideband regions and the signal region are not fully overlapped. To better characterize the backgrounds near the $M(K^+\Sigma^{*0})$ threshold, an energy shift is applied to the sideband events so that the invariant mass distribution $M(\Lambda\pi^0)$ is made to match the $\Sigma^{*0}$ signal region.  The correction formula is
\begin{equation}
\begin{split}
E_{\Sigma^{*0},\text{shift}} = \sqrt{(m_{\Lambda \pi^0} + \Delta_m)^2c^4 + (\vec{p}_{\Lambda} + \vec{p}_{\pi^0})^2c^2}, \\
M_{K^+\Sigma^{*0}} =
\sqrt{\frac{(E_{K^+} + E_{\Sigma^{*0},\text{shift}})^2}{c^4} - \frac{(\vec{p}_{K^+} + \vec{p}_{\Sigma^{*0}})^2}{c^2}},
\end{split}
\end{equation}
where $\Delta_m$ denotes the mass difference between the $\Sigma^{*0}$
signal region and sideband regions: $\Delta_{m,\text{left}} = 0.105\
\text{GeV}/c^2$ for the left sideband,  and $\Delta_{m,\text{right}} =
-0.105\ \text{GeV}/c^2$ for the right sideband. 
The corrected $\Sigma^{*0}$ sidebands are used to model the non-$\Sigma^{*0}$ backgrounds. As shown in Fig.~\ref{fig_shift1}, the mass-spectrum threshold in the corrected sideband region is consistent with that in the signal region.

 An unbinned maximum-likelihood fit is performed to the $M(K^+\Sigma^{*0})$ invariant mass distribution in the range $[1.85, 2.16]\ \text{GeV}/c^2$. Non-$\Sigma^{*0}$ backgrounds are estimated using the corrected $\Sigma^{*0}$ sidebands, with RooKeysPdf~\cite{Cranmer:2000du} employed in the fit and the number of background events fixed to the number of events in the sideband. Non-resonant backgrounds are described by the ARGUS function, where the threshold parameter is fixed, and all other parameters are free. The $N_{x}(1875)$ signal shape is given by $\epsilon(M_{K^+\Sigma^{*0}}) \times F_{\text{BW}} \otimes R(M_{K^+\Sigma^{*0}})$. Here, $\epsilon(M_{K^+\Sigma^{*0}})$ is the $M_{K^+\Sigma^{*0}}$ efficiency curve described by RooKeysPdf~\cite{Cranmer:2000du}, estimated from signal MC sample. $R(M_{K^+\Sigma^{*0}})$ is a double Gaussian function describing the detector resolution, and its parameters are derived from fits to signal MC sample. $F_{\text{BW}}$ is an S-wave BW function with a running width, which, in natural units $\hbar = c = 1$, is defined as
\begin{equation}
\begin{split}
F_{\text{BW}} = \left| \frac{1}{M_{N_{x}(1875)}^2 - s - iM_{N_{x}(1875)}\Gamma(s)} \right|^2, \\
\Gamma(s) = \Gamma_{N_{x}(1875)} \left( \frac{M_{N_{x}(1875)}}{\sqrt{s}} \right)^2 \left( \frac{q}{q_0} \right)^{2L + 1},L = 0,
\end{split}
\label{eq:FBW}
\end{equation}
where $M_{N_{x}(1875)}$ and $\Gamma_{N_{x}(1875)}$ are the mass and width of $N_{x}(1875)$ from the PDG~\cite{ParticleDataGroup:2024cfk}, fixed in the fit, while its quantum numbers are quoted from Ref.~\cite{He:2017aps}; $q$ is the momentum of $K^+$ or $\Sigma^{*0}$ in the rest frame of the resonance, and $q_0$ is the value of $q$ when $s = M_{N_{x}(1875)}^2$.

\begin{figure}
\centering
\includegraphics[width=0.33\textwidth]{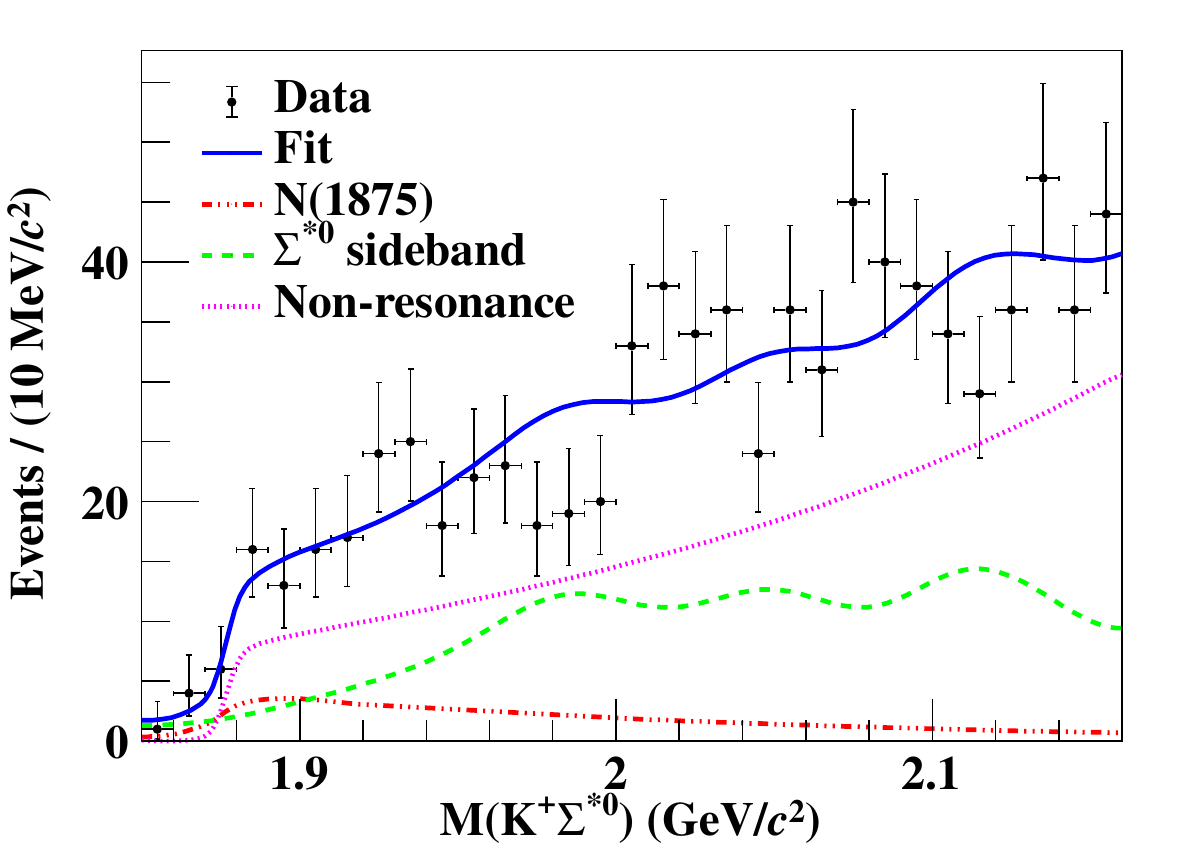}
    \caption{Fit to the $ K^+\Sigma^{*0} $ invariant mass
      spectrum. The dots with error bars are data; the red dot-dashed,
      green dashed and pink dotted lines represent the signal $N_{x}(1875)$,  non-$\Sigma^{*0}$ backgrounds and non-resonant backgrounds, respectively.}\label{fit1875}
\end{figure}

The result of the unbinned maximum-likelihood fit to $M(K^+\Sigma^{*0})$ is shown in Fig.~\ref{fit1875}. Since no significant $N_{x}(1875)$ signal is
observed, a Bayesian method~\cite{Rover:2011zq} is used to obtain the
upper limit on the signal yield at the 90\% C.L.. To determine the upper limit, the distribution of normalised likelihood values for a series of expected signal yields is taken as the probability density function. The upper limit of yield at 90\% C.L., $N_{\text{UL}}$, is set such that 90\% of the probability density function area above zero yield lies between 0 and $N_{\text{UL}}$. 

\subsection{Search for $\psi(3686) \to\bar{p}N_{x}(2100)\to \bar{p} K^{*0} \Sigma^{+}$}
\begin{figure}
\centering
\includegraphics[width=0.33\textwidth]{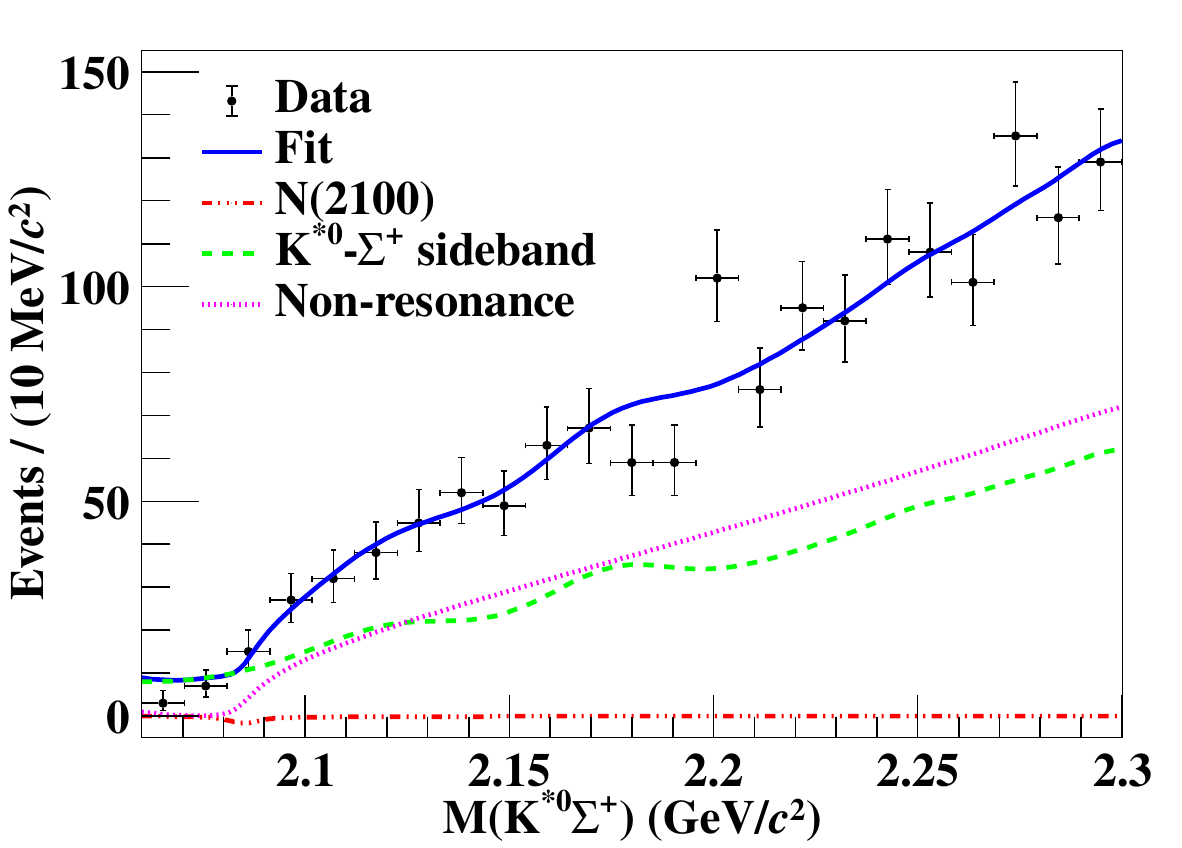}
    \caption{Fit to the $ K^{*0}\Sigma^+ $ invariant mass spectrum. The dots with error bars are data; the red dot-dashed,
      green dashed and pink dotted lines represent the signal $ N_{x}(2100) $,  non-$K^{*0}$–$\Sigma^{+}$ backgrounds and non-resonant backgrounds, respectively.}\label{fit2100}
\end{figure}

 Following the approach detailed in Sec.~\ref{sec:search1875}, an unbinned maximum-likelihood fit is performed to the $M(K^{*0}\Sigma^{+})$ invariant mass distribution in the range $[2.06, 2.30]\ \text{GeV}/c^2$. The backgrounds are found to fall into two categories: non-$K^{*0}$–$\Sigma^{+}$ and non-resonant. Non-$K^{*0}$–$\Sigma^{+}$ backgrounds are estimated using the corrected $K^{*0}$–$\Sigma^{+}$ sidebands. Non-resonant backgrounds are described by the ARGUS function. The signal shape model is the same as in Sec.~\ref{sec:search1875}. The fit result is shown in Fig.~\ref{fit2100}. Since no significant $N_{x}(2100)$ signal is observed, the Bayesian method~\cite{Rover:2011zq} is employed to set an upper limit on the signal yield at the 90\% C.L..

\begin{figure}
\centering
\includegraphics[width=0.33\textwidth]{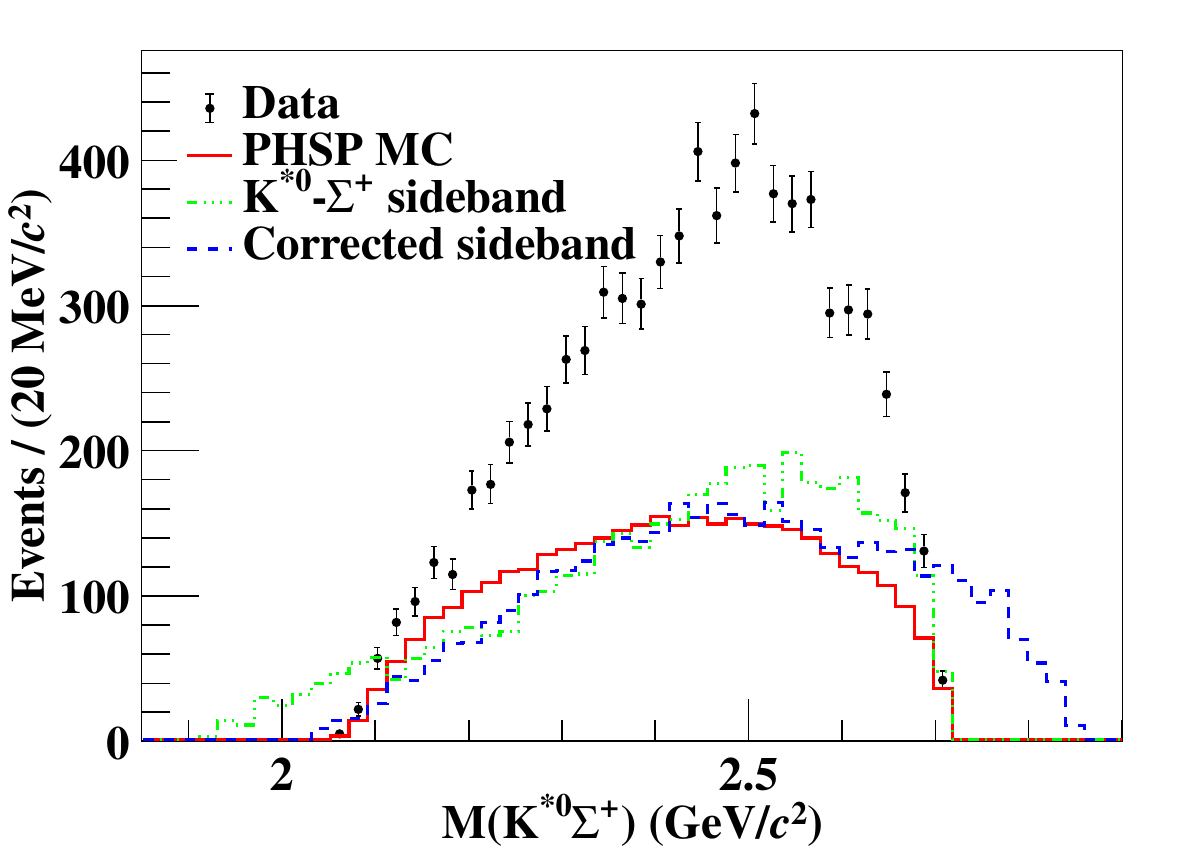}
    \caption{Invariant mass distribution of $K^{*0}\Sigma^{+}$. The dots with error bars are data; the red solid, green dot-dashed and blue
      dashed lines represent the PHSP MC sample, $K^{*0}$–$\Sigma^{+}$ sidebands and corrected sidebands, respectively. The PHSP MC is normalized to background-subtracted data events.}\label{fig_shift2}
\end{figure}

\subsection{Systematic uncertainties}
The systematic uncertainties are divided into two categories.
The first category consists of the multiplicative systematic uncertainties related to event selection, summarized in Table~\ref{tab_sysUBr}. The second category arises from the fit. We repeat the fitting procedure with different values of the signal mass and width, fitting ranges, sideband regions and non-resonant background shapes, and select the maximum upper limit among these cases.
\begin{table}
    \centering
    \caption{Systematic uncertainties (in \%) on the measurements of the upper limits of BFs (multiplicative).}
    \label{tab_sysUBr}
    \begin{tabular}{c|c|c}
        \hline\hline
        Source & \makecell{ $\psi(3686)\to$ \\ $\bar{p}N_{x}(1875)\to$ \\
        $\bar{p}K^+\Sigma^{*0}$} & \makecell{$\psi(3686)\to$\\
        $\bar{p}N_{x}(2100)\to$ \\
        $\bar{p}K^{*0}\Sigma^+$} \\
        \hline
        $N_{\psi(3686)}$          & 0.5    & 0.5    \\
        $\Lambda$ reconstruction  & 2.0    & -     \\
        $\Sigma^+$ reconstruction & -     & 1.9    \\
        MDC tracking              & 2.0    & 3.0    \\
        PID                       & 2.0    & 3.0    \\
        Photon detection          & 2.0    & -    \\
        \makecell{Quoted BFs }                & 1.9    & 0.6    \\
        Kinematic fit             & 0.9    & 0.5    \\ 
        \hline
        Total                     & 4.5    & 4.7    \\
        \hline\hline 
    \end{tabular}
\end{table}

To obtain a conservative upper limit on the BFs, we convolve the likelihood with a Gaussian function to incorporate systematic uncertainties. The resulting  smeared likelihood $L'(n')$ is given by
\begin{equation}
    \label{eq_Ln}
    L'(n')\varpropto \int_{0}^{1} L\left( n \frac{\epsilon}{\epsilon_0} \right) \exp\left[ \frac{-(\epsilon - \epsilon_0)^2}{2\sigma_\epsilon^2} \right]d\epsilon,
\end{equation}
where $L(n)$ denotes the Gaussian likelihood as a function of the signal yield $n$; $\epsilon_0$ is the detection efficiency estimated from signal MC sample; and $\sigma_\epsilon$ is the multiplicative systematic uncertainty. Using this smeared likelihood, the 90\% C.L. upper limits on the signal yields, $N_{\text{UL}}$, are $108.5$ and $17.1$ for $N_{x}(1875)$ and $N_{x}(2100)$, respectively.

The upper limits on the BFs for the decays $\psi(3686) \to\bar{p}N_{x}(1875)\to \bar{p} K^{+} \Sigma^{*0}$ and $\psi(3686) \to\bar{p}N_{x}(2100)\to \bar{p} K^{*0} \Sigma^{+}$ are calculated by
\begin{equation}
\mathcal{B} < \frac{N_{\text{UL}}}{N_{\psi(3686)} \cdot \prod_{i} \mathcal{B}_{i} \cdot \epsilon},
\label{eq:br_ul_unified}
\end{equation}
where $\prod_{i} \mathcal{B}_{i}$ denotes the product of BFs of subsequent decay channels, with values taken from the PDG~\cite{ParticleDataGroup:2024cfk}; $\epsilon$ is the efficiency estimated from signal MC sample.

For the decay $\psi(3686) \to\bar{p}N_{x}(1875)\to \bar{p} K^{+} \Sigma^{*0}$,
    $\prod_{i} \mathcal{B}_{i}=\mathcal{B}(\Sigma^{*0} \to \Lambda \pi^0) \cdot \mathcal{B}(\Lambda \to p \pi^-) \cdot \mathcal{B}(\pi^0 \to \gamma\gamma)$ and $\epsilon=(4.63 \pm 0.02)\%$. The upper limit of $\mathcal{B}(\psi(3686) \to\bar{p}N_{x}(1875)\to \bar{p} K^{+} \Sigma^{*0})<1.58 \times 10^{-6}$ is obtained at the 90\% C.L..
    For the decay $\psi(3686) \to\bar{p}N_{x}(2100)\to \bar{p} K^{*0} \Sigma^{+}$,
    $\prod_{i} \mathcal{B}_{i}=\mathcal{B}(K^{*0} \to K^+ \pi^-) \cdot \mathcal{B}(\Sigma^+ \to p \pi^0) \cdot \mathcal{B}(\pi^0 \to \gamma\gamma)$ and $\epsilon=(6.06 \pm 0.02)\%$. The upper limit of $\mathcal{B}(\psi(3686) \to\bar{p}N_{x}(2100)\to \bar{p} K^{*0} \Sigma^{+})<3.09 \times 10^{-7}$ is obtained at the 90\% C.L..

\section{Summary}

Based on a sample of $(2712.4 \pm 14.3) \times 10^6$ $\psi(3686)$ events collected with the BESIII detector, the decays $\psi(3686)\to\bar{p} K^{+} \Sigma^{*0}+c.c.$ and $\psi(3686)\to\bar{p} K^{*0} \Sigma^{+}+c.c.$ are observed for the first time and their BFs are measured to be $\mathcal{B}(\psi(3686)\to\bar{p}K^+\Sigma^{*0}+c.c.)=(3.94\pm0.15_{\rm stat.}\pm0.24_{\rm syst.})\times10^{-5}$ and $\mathcal{B}(\psi(3686)\to\bar{p}K^{*0}\Sigma^++c.c.)=(4.93\pm0.14_{\rm stat.}\pm0.30_{\rm syst.})\times10^{-5}$. No significant signals for the intermediate state $N_{x}(1875)$ or $N_{x}(2100)$ are observed in the invariant mass spectra of $K^{+}\Sigma^{*0}$ and $K^{*0}\Sigma^{+}$, respectively. Upper limits on the BFs at the 90\% C.L. are set to be $\mathcal{B}(\psi(3686)\to\bar{p}N_{x}(1875)\to\bar{p}K^+\Sigma^{*0})<1.58\times10^{-6}$ and $\mathcal{B}(\psi(3686)\to\bar{p}N_{x}(2100)\to\bar{p}K^{*0}\Sigma^+)<3.09\times10^{-7}$. These results provide experimental input and constraints for relevant theoretical studies of hidden-strange pentaquark states.

\section*{Acknowledgment}
\input{acknowledgement_2026-05-08}

\section*{BESIII Collaboration}
\input{authorlist_2026-05-08}

\bibliographystyle{elsarticle-num} 
\bibliography{reference}
\end{document}

%% file: acknowledgement_2026-05-08.tex
The BESIII Collaboration thanks the staff of BEPCII (https://cstr.cn/31109.02.BEPC) and the IHEP computing center for their strong support. This work is supported in part by National Key R\&D Program of China under Contracts Nos. 2025YFA1613900, 2023YFA1606000, 2023YFA1606704; National Natural Science Foundation of China (NSFC) under Contracts Nos. 11635010, 11935015, 11935016, 11935018, 12025502, 12035009, 12035013, 12061131003, 12192260, 12192261, 12192262, 12192263, 12192264, 12192265, 12221005, 12225509, 12235017, 12342502, 12361141819, 12535005; the Chinese Academy of Sciences (CAS) Large-Scale Scientific Facility Program; the Strategic Priority Research Program of Chinese Academy of Sciences under Contract No. XDA0480600; CAS under Contract No. YSBR-101; 100 Talents Program of CAS; The Institute of Nuclear and Particle Physics (INPAC) and Shanghai Key Laboratory for Particle Physics and Cosmology; Agencia Nacional de Investigaci\'on y Desarrollo de Chile (ANID), Chile under Contract No. ANID CCTVal CIA250027; ERC under Contract No. 758462; German Research Foundation DFG under Contract No. FOR5327; Istituto Nazionale di Fisica Nucleare, Italy; Knut and Alice Wallenberg Foundation under Contracts Nos. 2021.0174, 2021.0299, 2023.0315; Ministry of Development of Turkey under Contract No. DPT2006K-120470; National Research Foundation of Korea under Contract No. RS-2026-25486791; National Science and Technology fund of Mongolia; Polish National Science Centre under Contract No. 2024/53/B/ST2/00975; STFC (United Kingdom); Swedish Research Council under Contract No. 2019.04595; U. S. Department of Energy under Contract No. DE-FG02-05ER41374

%% file: authorlist_2026-05-08.tex
M.~Ablikim$^{1}$\BESIIIorcid{0000-0002-3935-619X},
M.~N.~Achasov$^{4,c}$\BESIIIorcid{0000-0002-9400-8622},
P.~Adlarson$^{84}$\BESIIIorcid{0000-0001-6280-3851},
X.~C.~Ai$^{90}$\BESIIIorcid{0000-0003-3856-2415},
C.~S.~Akondi$^{32A,32B}$\BESIIIorcid{0000-0001-6303-5217},
R.~Aliberti$^{40}$\BESIIIorcid{0000-0003-3500-4012},
A.~Amoroso$^{83A,83C}$\BESIIIorcid{0000-0002-3095-8610},
Q.~An$^{79,66,\dagger}$,
Y.~H.~An$^{90}$\BESIIIorcid{0009-0008-3419-0849},
M.~S.~Anderson$^{40}$\BESIIIorcid{0009-0008-1550-2632}
% ...
\\
\vspace{0.2cm}
(BESIII Collaboration)\\
\vspace{0.2cm} {\it
$^{1}$ Institute of High Energy Physics, Beijing 100049, People's Republic of China\\
% ...
$^{4}$ Budker Institute of Nuclear Physics SB RAS (BINP), Novosibirsk 630090, Russia\\
% ...
$^{32}$ INFN Sezione di Ferrara, (A)INFN Sezione di Ferrara, I-44122, Ferrara, Italy; (B)University of Ferrara, I-44122, Ferrara, Italy\\
% ...
$^{40}$ Johannes Gutenberg University of Mainz, Johann-Joachim-Becher-Weg 45, D-55099 Mainz, Germany\\
% ...
$^{66}$ State Key Laboratory of Particle Detection and Electronics, Beijing 100049, Hefei 230026, People's Republic of China\\
% ...
$^{79}$ University of Science and Technology of China, Hefei 230026, People's Republic of China\\
% ...
$^{83}$ University of Turin and INFN, (A)University of Turin, I-10125, Turin, Italy; (B)University of Eastern Piedmont, I-15121, Alessandria, Italy; (C)INFN, I-10125, Turin, Italy\\
$^{84}$ Uppsala University, Box 516, SE-75120 Uppsala, Sweden\\
% ...
$^{90}$ Zhengzhou University, Zhengzhou 450001, People's Republic of China\\

\vspace{0.2cm}
$^{\dagger}$ Deceased\\
% ...
$^{c}$ Also at the Novosibirsk State University, Novosibirsk, 630090, Russia\\
% ...

}